# Developing Open Data Models for Linguistic Field Data

## Baden Hughes

*Department of Computer Science and Software Engineering*
*University of Melbourne, Victoria 3010, Australia*
*badenh@cs.mu.oz.au*

## 1. Introduction

The UQ Flint Archive houses the field notes and elicitation recordings made by Elwyn Flint in the 1950's and 1960's during extensive linguistic survey work across Queensland, Australia. The linguistic fieldwork documents 54 Australian Aboriginal languages, of which approximately half are now extinct, and the remainder are in various stages of endangerment. The field notes from this survey amount to approximately 900 separate documents including elicitation lists, phonological sketches, grammatical notes and transcriptions and is all in handwritten paper format. Corresponding audio recordings, originally made on reel to reel tape, have been converted to more modern formats, and comprise a collection of CDROM media.

The primary aim of the digitization project is to provide a web-based portal where Aboriginal languages can be explored, and through which new research can be facilitated. Work on the digitization of the UQ Flint Archive has been carried out since 1996, using various technologies and approaches. Recently, significant progress has been made in the analysis of the technical requirements and an overall strategy for the completion of the project.

The process of digitizing the contents of the UQ Flint Archive provides a number of interesting challenges in the context of EMELD. Firstly, all of the linguistic data is for languages which are either endangered or extinct, and as such forms a valuable ethnographic repository. Secondly, the physical format of the data is itself in danger of decline, and as such digitization is an important preservation task in the short to medium term. Thirdly, the adoption of open standards for the encoding and presentation of text and audio data for linguistic field data, whilst enabling preservation, represents a new field of research in itself where best practice has yet to be formalised. Fourthly, the provision of this linguistic data online as a new data source for future research introduces concerns of data portability and longevity.

This paper will outline the origins of the data model, the content creation components, presentation forms based on the data model, data capture tools and media conversion components. It will also address some of the larger questions regarding the digitization and annotation of linguistic field work based on experience gained through work with the Flint Archive contents.

## 2. Project Objectives

To place the development of open data models in context, it is useful to review the high level objectives of the overall Flint Archive digitization project. In general, project objectives were established with with reference to the criteria provided by Bird and Simons (2002).

A summary of the relevant objectives follows (Hughes, 2002:5) :

- To develop a searchable electronic catalogue, using an open standard such as XML
- To provide this catalogue on the Web using a custom built web server interface or a third party provider interface such as that provided by OLAC.
- To define XML representations for data based on existing templates
- To develop a series of tools to allow the creation of new documents within these templates (that is, to enable the keying in of source materials directly into a usable format)
- To develop a conversion toolkit for existing electronic data

- To provide the ability to convert core data sources into numerous usable electronic format for linguistic researchers to download directly
- To present each set of language in a standardized manner on the web, possibly modeled after language descriptions in print, and including integrated media sources (particularly audio)

In progressing toward these objectives, a number of different processes and outputs of relevant to EMELD have been defined. We will first examine the data types and sources, and then consider the proposed solutions for an open data model for linguistic field data in this context.

## *3. Original Data Types and Sources*

Within the Flint Archive there are number of different data types and sources. At a high level there exists metadata, text, audio and images, a long term storage format, and presentation requirements. We will discuss each of these in turn.

Initial work on contents of the Flint Archive was institutionally funded in the mid 1990's and resulted in a number of significant outputs including a media preservation strategy, a basic catalogue, and some digitization of audio. While the amount of material digitized through this process is small compared to the overall archive contents, it nevertheless provided a basic structure for future work, and as such warrants review.

## 3.1 Metadata

Previous work (Laughren et al, 1996) has surveyed the archive contents and published a basic HTML based catalogue of the contents of the archive based on Template 1 (see Appendix 1). This is the template used to encode cross-referencing information and recording details, which we can view essentially being language resource metadata. A separate representation of the catalogue is included within the UQ Library's information system and is accessible as raw MARC records.

## 3.2 Textual Materials

The textual materials within the archive consist some 900 separate documents, of which the vast majority are (badly) handwritten records, with some standardized forms which largely contain metadata. A small amount of material is typed, perhaps suitable for OCR, however the majority is not suitable for any automated processing. Previous work has resulted in materials from 2 languages being available electronically – elicitation (lexical) material encoded according to Template 2 (see Appendix 2), and textual material (primary text) material encoded according to Template 3 (see Appendix 3). Of this electronic material, materials from one language (Yanyula) has been modified into statically linked HTML (Laughren, Keith and Yuen, 1999), and from another language (Garrwa) has dynamically linked based on an XML format (Laughren, Keith and Hughes 2002). No other field linguistic data has been digitized. There remains a significant task to digitize all of the textual material within the archive – in the meantime the physical archive holdings are in controlled environmental storage.

## 3.3 Audio

Previous work has commenced the digitization of audio resources within the archive. The original format for audio was on a reel to reel tape, which has been converted into a digital format and archived on CDROM for long term storage.

## 3.4 Images

During fieldwork, a large number of photographs of ethnographic significance were also collected. These are generally in a poor physical state, although exploratory efforts at scanning these images for posterity have been undertaken.

## *4. Target Structures and Formats*

Having considered the contents of the archive, we now examine the technical objectives, namely the structural representations and formats for both presentation and long term storage.

## 4.1 Structure

Earlier work has resulted in the definition of three templates used to structure and encode data. While these provide a good depth of structure for text, an analysis of the "lexicon" (Template 2) and the "primary text" Template 3) indicates a very high degree of similarity between the two. Inclusions in Template 2 which differentiate it are the "Other Recordings" and "Semantic or Syntactic Domain" elements, whilst the inclusion in Template 3 which differentiates it is the inclusion of the element "Speaker".

This similarity motivates a unification approach to developing a single template that can be used to encode both lexicon and primary text linguistic data types. By adopting an XML encoding, the differences between the two types can easily be included as an element property or schamatised, hence bringing efficiency to future data entry and unifying the two separate templates into a single data representation.

The original Template 1, which basically structures metadata, will be modified to be OLAC compliant with extensions for the Flint Archive specific content where necessary. Since the content of this particular template applies across a number of text and audio sources, a single XML representation will be developed but it is envisaged that this will commonly be included with other data.

## 4.2 Formats

We now discuss the target data formats for metadata, text, and audio which have been agreed upon in the context of the Flint Archive.

### 4.2.1 Metadata
The target metadata format is the OLAC metadata set, with an automated conversion of the existing catalogue to a valid XML document as an interim step to full OLAC implementation. Other systems, such as library information systems, require MARC format, which can be achieved by an OLAC to Dublin Core crosswalk and a subsequent Dublin Core to MARC crosswalk. As a part of the process of converting the catalogue, a review of classification was also undertaken.

### 4.2.2 Text
The target text format is UTF-8. This is based on the default requirement for XML compatibility. Given that Australian languages often use practical orthographies, it is unlikely that any characters not included in Unicode will be required, and as such, adopting a non Unicode format is inefficient.

Additionally, a longer term project is to scan and image the original handwritten manuscripts to create an image archive. Images generated from such a process would conform to the standards specified in section 4.2.4.

### 4.2.3 Audio
The target audio format is divided between a long term high definition format, and an efficient, flexible format for web based publication. As such, 44kHz WAV format has been selected as the long term format; whilst 16kHz MP3 format has been selected as the web targeted format (which balances file size with quality and utility). A library of MP3 audio files will be compiled onto CD or DVD at the completion of the project.

### 4.2.4 Images
The target image format is similarly divided between a long term high definition format and a format motivated by a web publishing agenda. In this case, images were scanned at 1200dpi into TIFF format, and then rasterised and converted into JPEG format for immediate use.

## *5. Open Data Models*

We now turn to a discussion of the processes involved in developing the open data models, discussing the creation of an XML representation for data, input tools, conversion and rendering tools, and finally publishing outputs.

## 5.1 XML Representation

Based on the unification work for the lexicon and primary text templates, an XML data model has been developed, with a corresponding DTD. Although the data entry environment enforces structural constraints, the availability of a DTD allows separate validation of the XML document on demand. A valid XML document is significantly easier to manipulate programmatically in general, and specifically in terms of presentation output. We provide a sample XML fragment in Appendix 4.

As can be observed, the XML data representation is essentially interlinear, although only at the phrase level. The use of element codes and controlled types allows flexibility with regard to description of the textual content.

## 5.2 Input Tools

Given the nature of the linguistic data in the archive (namely, handwritten papers requiring keying in), the selection of a data input tool or tools is of significant importance in enabling the ongoing work on the archive. Experiments have been conducted using three different interfaces for the keying of language data based on the new data representation.

The first interface choice is perhaps the simplest. Any plain text editor can be used to enter data in a canonical format, and subsequent conversion of the field ordered standard format (FOSF, or "back-slashed") files can be enabled. There are several issues with this approach, namely that XML requires well-formed elements (with start and end tags), and the FOSF files only explicitly provide start tags, and thus end tags need to be inferred. However, these plain text files are easily created on a range of platforms and using a range of basic word processing tools, which does make this somewhat more attractive as an approach to be adopted.

The second interface choice is to develop a customized data input environment which supports the project. Due to funding and technical constraints, this was not viewed as an efficient direction to take. This is reinforced by the absence of formally specified general models for linguistic field data, and as such, any development in this effort would serve to add yet another data entry environment to the variety of those already available.

The third, and the approach ultimately selected is to use a Microsoft Excel template, which has the distinct advantage of allowing two dimensional structures (columns and rows), whilst retaining ease of export functions. Since the editing functionality required is only basic, it is also possible to use the Excel templates on Windows, Macintosh and Unix systems (the latter through OpenOffice). An additional advantage is that constraints can be embedded in cell logic to assist data entry. Perhaps the most notable is the ability to cross reference resources such as audio and text without "embedding" them into the actual document. There are some minor disadvantages, namely that it is safe to assume that not all potential date entry operators are literate with Excel and that the temptation to "enhance" by WYSIWYG formatting is still present, but both of these are to a certain extent able to be compensated for. This interesting choice is discussed further in section 6 below.

## 5.3 Format and Conversion Tools

We now consider the tools used in the ongoing process of digitizing the Flint Archive contents. The primary functions of these tools are to convert data (text, audio and images) into long term storage and presentation formats as defined earlier in 4.2.

### 5.3.1 Text

The tools selected for text manipulation were a custom data conversion utility named XL2XML (Hughes 2001), and UltraEdit (IDM Computer Solutions, 2003). XL2XML is a data conversion application written in Visual Basic 6 which runs on Windows. XL2XML allows the export of two dimensional data structures from Microsoft Excel into an XML document – there are a number of commercial tools available which essentially perform the same function. This tool is used to convert the data entered in Excel into the underlying XML representation. UltraEdit is a lightweight programming editor for Windows. It provides a wide range of text utility functions, integration with a number of programming environments, and has an open and extendable scripting function. This tool is used to make minor adjustments to the resulting XML files.

### 5.3 2 Audio

The tools selected for audio manipulation were the increasingly common Goldwave Digital Audio Editor, (Goldwave Inc, 2003), and an associated plugin, Razor LAME (The LAME Project, 2003). Goldwave is a fully featured Windows based audio editing package which supports a large number of formats, and includes a full suite of remastering controls along with a batch mode interface. Razor LAME is an open source MP3 conversion toolkit which allows Goldwave to handle advanced MP3 encoding and decoding functions. These two tools in combination were used to batch convert the WAV format audio files into MP3 format, and to segment the audio files.

### 5.3.3 Images

The tool selected for image manipulation was Graphic Workshop Professional (Alchemy Mindworks, 2003). Graphic Workshop Professional is Windows based an industrial quality, shareware priced image conversion toolkit, supporting a large number of image formats (including moving image formats), and a wide variety of image manipulation techniques. A significant feature of this program is its remotely instantiable batch mode which makes processing large numbers of source images much more efficient. This tool was used to convert *en masse* the scanned images contained within the archive.

## 5.4 Publishing Outputs

An important component of the Flint Archive digitization project is to enable researchers to easily interact with the linguistic field data which has until recently been inaccessible. A major consideration in the process of publishing this data to retain longevity and flexibility whilst providing formats of interest to linguists which in themselves may be re-purposed. An additional factor is the project vision that end users would be able to add analysis to the data through some kind of a collaborative annotation process.

In addressing the publishing output question a number of desirable outputs were identified and are here extracted from (Hughes, 2002:12). Presentation types of interest include lexicons (dictionary style with configurable Target:Source options), word lists (again with Target:Source options), optional display formats (based on granularity of interlinear), audio file download, and PDF generation. In all of these cases, the use of open source software in enabling these outputs is viewed as desirable.

In order to meet these requirements, an approach based on XSL has been adopted. XSL allows a single underlying data source to be rendered in a range of different presentation formats depending on requirements, and is a companion to the XML standard. Through the use of XSL, the same Aboriginal language data can be presented in a format resembling a traditional print description, or converted into one of the linguistic outputs described earlier. XSL also allows linkage to local and non-local resources such as audio files, and the integration of multiple data sources into a single presentation format.

## 6. EMELD Context

Through the process of developing open data models for linguistic field data in the context of the Flint Archive a number of correspondences to various goals of the EMELD project have emerged. It is therefore useful to evaluate the model based on practical experience in working with linguistic data encoded according to these standards. In this process we will identify a number of strengths and weaknesses of this model and propose some wider EMELD implications. In particular, the Flint Archive process has discovered and addressed issues such as how to align text and transcriptions, how to align and link text and audio, how to structure interlinear corpora and how to integrate data models for lexical and textual information.

We will next turn to a discussion of some specific references and general cases within the EMELD context.

In relation to metadata for language resources (EMELD Proposal 3.2), the Flint Archive project has addressed these issues by adopting the OLAC metadata set as the basis for language resource descriptions and by adopting the corresponding OLAC extensions for language identification, linguistic field and linguistic type as the basic vocabulary for typological metadata. Given the nature of the project, formal implementation of a metadata collection mechanism in a linguistic fieldwork context has not been required.

In relation to markup for language resources (EMELD Proposal 3.3), the data model proposed addresses both the requirements for encoding glossed text and lexical entries through a single flexible interlinear format. In this model, a lexical entry is a minimal case of an interlinear text, a concept which provides leverage for other data models (such as a purely lexical representation). Owing to the flexibility of the underlying structure, any XML editing application can then be used to modify the corpus or the textual materials. Ontologies for the description of each constituent of a lexical item or text are drawn from higher level controlled vocabularies of linguistic terminology.

In relation to data formats and software tools (EMELD Proposal 3.4), the Flint Archive project has adopted an XML-based data structure that is converted to a human readable format for display. By adopting an open, extensible underlying data structure, existing electronic data can be converted efficiently, whilst new data can be entered in a variety of environments and tools. In particular, we allow non-linguistic applications to be utilized for the data entry task, and the resulting intermediate forms are easily converted to the underlying data model.

Next we will discuss general recommendations stemming from our work on the Flint Archive data model.

As mentioned earlier, the data model developed is fundamentally interlinear, reflecting earlier work on digitizing the archive contents which adopted this modality. The interlinear implementation exhibits a high degree of flexibility in its ability to include multiple, extendable levels and types of glossing with minimal impact on the overall text format. Whilst this is beneficial, the model does not adequately address issues relating to the alignment of glossing with original text except at the phrase level in a primary text modality, or at the word level in a lexical modality. This shortcoming is tolerable in the Flint Archive context since alignment is not a high priority outcome and in many cases, alignment is subject to incomplete linguistic analysis of the language data which would in turn enable this particular feature. In the broader context of EMELD, we can identify a requirement that a general model for interlinear text in particular must support a high degree of granularity in the number and types of glosses, in addition to sufficient refinement to enable small linguistic unit level alignment (eg phoneme and morpheme, in addition to words and phrases).

Another point of criticism is that the data model, whilst based on open international standards, and designed for encoding data from a moderate number of languages (approx. 60), is constrained by its lack of exposure to linguistic data of other formats (essentially the Flint Archive data is heterogenous in terms of the type of linguistic information, primarily of OLAC Linguistic Type (Aristar Dry and Johnson, 2002) "Lexicon" and "Primary Text"). As such it may not adequately address issues which arise in different linguistic data collection contexts, for example data which contains a high proportion of multiple participant discourse. Implications for EMELD are that data models need to retain a high degree of flexibility in order to be useful in cross-linguistic and cross-linguistic data type contexts.

Many different methods of integrating time-series multimedia have been described by both computer scientists and linguists. This variety of approaches reflects the complexity of developing and promoting open models in this area. In addressing this issue, our data model provides only basic support for the integration of text and audio. By using a URL to identify the audio source and a basic time offset in the textual material, we provide only a minimally useful solution to this problem. While this is an inherent weakness, larger, better resourced projects have struggled to provide approachable solutions to this problem which are widely adopted and supported by common software tools. A particular criticism of our approach is that it inherits structure from the type of linguistic data present in the archive – audio is typically recordings of elicitation sessions based on a Capell word list (Capell, 1945). Variance from this style of audio would introduce new complexity for the basic time offset approach, especially in the context of continuous speech flows. Implications for EMELD are that any data model must include provision for the alignment of (and possibly embedding of references to) times series data across text, audio and video domains. Ideally such support should include provision for non-local resources, and potentially both a high-definition resource and an efficient web-based version.

One strength of the approach taken with the Flint Archive holdings is the adoption of commonly available, off the shelf commercial applications for data entry. By utilizing a data model built on open international standards, we increase the likelihood that existing editing applications may be adopted for editing linguistic field data.This differentiates the Flint project from other similar digitization projects where customized software is often written or adapted to enable data capture according to different data models. As discussed earlier, even commonly used non-linguistic applications (such as Microsoft Excel) can be used to enter structured data which is then converted into our open standard. In this case the emphasis is that an adequate path must be identified and supported in order to ensure that where selected, proprietary tools do not lock data away in unfriendly formats. Implications for EMELD include consideration that not only should "best practice" tools be developed to support EMELD standards, but that information needs to be collated and disseminated regarding possible adaptions of other data manipulation environments to support EMELD standards either directly or indirectly.

Working with linguistic field data in the Flint Archive context differs from working with other linguistic field data. This is by virtue of the fact that the process is fundamentally one of retrofitting structures to existing archival materials rather than directly enabling field data capture. As a consequence, data models and tools are driven by different requirements to those which enable the documentary field linguist to collate data. In the context of EMELD this raises another important requirement, namely, that standards and tools must be compatible across different modes of working with linguistic field data, and not simply account for one or another type of practice. Ideally, data models would be applicable across both archival and field linguistic work environments, whilst data capture tools may vary depending on focus. The uniformity of an underlying data format is of significant benefit, at a minimum this provides a leverage point between contexts.

## 7. Conclusion

The process of digitizing the contents of the UQ Flint Archive has provided a number of interesting findings in the broader context of EMELD and the agenda of digitizing lexical information and linguistic field data. It has been shown how a combination of open standards, open source and proprietary tools can be appropriately leveraged to provide an overall approach to digitization of such materials, whilst retaining a high degree of efficiency, flexibility and scalability. Reflections on this process in the context of EMELD has provided a range of new requirements and recommendations for general models and standards for the digitization of linguistic field data.

# *Appendix 1*

## Flint Archive Template 1 (Metadata) in Field Ordered Standard Format with Descriptions.

| SF Marker | Name | Comment |
|---|---|---|
| \dn | Document Number | Document Number = Box Number + Document Number + Subdocument Number. |
| \cr | Cross References | Document Number of associated documents within the Archive. |
| \fl | Flint's Tape Log Number | Reference number of tape in Flint's tape log.. |
| \an | AIATSIS Acquisition Number | AIATSIS acquisition number of the audio tape as part of the Flint Collection in the AIATSIS sound archives. |
| \cm | AIATSIS Tape Reference Code | AIATSIS reference code for the audio tape. |
| \cd | AIATSIS CDROM Reference Code | AIATSIS reference code for digitized audio on CDROM. Usually same as AIATSIS tape reference. |
| \gp | Group Number | Number assigned by Flint to informant group. |
| \lg | Language Name | Name of language being recorded (Flint's spelling). |
| \loc | Recording Location | Location where recording took place. |
| \da | Recording Date | Date and time when recording took place. |
| \o | Observer | Researcher eliciting data for recording. |
| \snc | Speaker Name and Code | Informant providing language data. Optionally repeated for more than one informant. If the language material is primary text (rather than lexicon), so that there is alternation between speakers, each speaker is assigned a code (generally the first letter of their name), which is then used to specify the speaker in the text (in the \sp field of Template 3). |

## *Appendix 2*

## Flint Archive Template 2 (Lexicon) in Field Ordered Standard Format with Descriptions

| SF Marker | Name | Comment |
|---|---|---|
| \ft | Flint's Transcription | Flint's Transcription of Aboriginal words. |
| \or | Other Recordings | Indicates where multiple utterances of a single entry, where they have been recorded. For some entries, the informant has given a word or phrase several times, and, where appropriate, these repetitions have been recorded. |
| \ncr | Comments on Recording | Comments regarding special features of the recording, or regarding extra information given in the recording. It is used, for instance, when the informant has given a phrase in parts (rather than as a whole) to allow Flint time to write the words down. |
| \sd | Semantic or Syntactic Domain | Tthe semantic or syntactic category to which a word or clause has been assigned. |
| \ncft | Comments on Flint's Transcription | Comments regarding Flint's transcription of the word or words. It is used, for instance, to comment upon the presence of diacritics, marks whose significance is not clear, and phonemic symbols which appear rarely in the data for a given language. |
| \fg | Flint's Gloss | This field accommodates the English word or words recorded by Flint which correspond directly to the words in Flint's transcription. In other words, it is a gloss rather than a free translation. |
| \ncfg | Comments on Flint's Gloss | Comments on Flint's gloss. |
| \fft | Flint's Free Translation | Flint's free English translation rather than a word-by-word English gloss. |
| \ncfft | Comments on Flint's Free Translation | Comments on regarding Flint's free translation. |
| \os | Other Sources | This field accommodates relevant entries from other sources eg formal grammars published post recording date. These entries are transliterated according to the same system as is used for Flint's transcription. |
| \ncos | Comments on Other Sources | Comments on other sources |
| \na | Analysis | Attempted syntactic, morphemeic or phonological analysis of Flint's transcription. |
| \ncna | Comments on Analysis | Comments explaining or qualifying the analysis. It is used, for instance, when more than one analysis appears to be possible or where a published source conflicts with Flint's analysis. |
| \ng | Gloss | This field accommodates a morpheme-by-morpheme gloss of the analysis given in \na. |
| \ncng | Comments on Gloss | This field accommodates comments regarding the morpheme-by-morpheme gloss. It is used, for instance, when we are assuming or hypothesising that a morpheme corresponds to a morpheme given in other sources, even when the two are not identical. Where there is a discrepancy between the gloss Flint has given and the meanings supplied by other sources, the gloss given by Flint is generally incorporated into \ng. |

## *Appendix 3*

## Flint Archive Template 3 (Primary Text) in Field Ordered Standard Format with Descriptions

| SF Marker | Name | Comment |
|---|---|---|
| \sp | Speaker | Code of speaker producing particular line of text. |
| \ft | Flint's Transcription | Flint's transliteration of Aboriginal words. |
| \fg | Flint's Gloss | Flint's gloss of Aboriginal words. |
| \ncr | Comments on Recording | Comments on notable features of the recording, or regarding extra information given in the recording. Eg when the informant has given a phrase in parts (rather than as a whole) to allow Flint time to write the words down. |
| \ncft | Comments on Flint's Transcription | Comments regarding Flint's transcription of the word or words. eg comment upon the presence of diacritics, marks whose significance is not clear, and phonemic symbols which appear rarely in the data for a given language. |
| \ncfg | Comments on Flint's Gloss | Comments on Flint's gloss. |
| \fft | Flint's Free Translation | Flint's free English translation rather than a word-by-word English gloss. |
| \ncfft | Comments on Flint's Free Translation | Comments regarding Flint's free translation. |
| \os | Other Sources | This field accommodates relevant entries from other sources eg formal grammars published post recording date. These entries are transliterated according to the same system as is used for Flint's transcription. |
| \ncos | Comments on Other Sources | Comments regarding other sources. |
| \na | Analysis | Attempted syntactic, morphemeic or phonological analysis of Flint's transcription. |
| \ncna | Comments on Analysis | Comments explaining or qualifying the analysis. It is used, for instance, when more than one analysis appears to be possible or where a published source conflicts with Flint's analysis. |
| \ng | Gloss | This field accommodates a morpheme-by-morpheme gloss of the analysis given in \na. |
| \ncng | Comments on Gloss | Comments regarding the morpheme-by-morpheme gloss. It is used, for instance, when we are assuming or hypothesising that a morpheme corresponds to a morpheme given in other sources, even when the two are not identical. Where there is a discrepancy between the gloss Flint has given and the meanings supplied by other sources, the gloss given by Flint is generally incorporated into this line. |

## Appendix 4

## Sample XML Fragment (Garrwa, from Laughren, Keith and Hughes, 2002).

```
<text type="lexicon">
<media format="audio/mp3" url="302b.mp3" comment=""/>
<speaker code="G">Stumpy George</speaker>
<lexical_item offset="0.00" syn-sem_domain code="adjective">
<flint_transcription>'barki"nani</flint_transcription
<flint_transcription_comment/>
<other_recordings/>
<recording_comment>Informant's pronunciation approaches "balkinan" </recording_comment>
<flint_gloss>bad</flint_gloss>
<flint_gloss_comment>English gloss on the recording: "bad; bad fellow; bad man" </
flint_gloss_comment>
<flint_free_translation/>
<flint_free_translation_comment/
<other_sources>
<source>Belfrage 1997:2: balki (adj) ("bad")</source>
<source> Osborne 1966:8:
  "balkin "nanama bujili
  no good that billy can
  That billy can's no good.
</source>
<source> In Osborne's (1966) data,"nani appears in three sentences as the modifier "that". </
source>
</other_sources>
<other_sources_comment>
The adverb nani ("like this, like that") does not seem to fit particularly well into the
present item in Flint's data. Analysing nani as a demonstrative pronoun or as a nominalising
suffix is a more attractive option (see "comments on gloss" field below).
</other_sources_comment>
<analysis code="Naomi_Keith">barki nani </analysis>
<analysis_comment/>
<gloss>bad (fellow) that one</gloss>
<gloss_comment>It is difficult to determine the exact meaning of nani:
It may be functioning as the demonstrative pronoun "that one" (translated "he") in a verbless
topic-comment clause (i.e. "He's a bad fellow"). This use of nani is not attested in Osborne's
(1966) data, where nani ("that") modifies a head noun in the three examples in which it
appears. Similarly, "nana ("that") also seems to function as a modifier in singular sentences.
It may be worth noting that there is some freedom in the use of another demonstrative pronoun,
nanama(n), in Osborne's data: nanama(n) occurs both as the head of noun phrases and as a
modifier of head nouns. Perhaps this flexibility is also a feature of nani. Alternatively, nani
may be functioning as a nominalising suffix, with a meaning like "-one" (yielding "bad one",
"good one" etc.). This option is attractive because it fits well with Flint's elicitation on
the recording: "bad; bad fellow; bad man".Without more data, it is difficult to determine the
function of nani here. </gloss_comment>
</lexical_item>
<text>
```